\documentclass[%
 reprint,
twocolumn, 
 amsmath,amssymb,
 aps,
prl,
superscriptaddress
]{revtex4-2}

\usepackage{graphicx}
\usepackage{dcolumn}
\usepackage{bm}
\usepackage[colorlinks=true, allcolors=blue]{hyperref}
\usepackage{braket}
\usepackage{bbding}
\usepackage{natbib}
\usepackage{gensymb}
\usepackage{balance}
\usepackage{makecell}
\usepackage{multirow}
\usepackage{longtable}
\usepackage{comment}

\begin{document}

\preprint{APS/123-QED}

\title{Electric-Switchable Chiral Magnons in ${\cal PT}$-Symmetric Antiferromagnets}

\author{Jinyang Ni}
\email{These authors contributed equally to this work}
\affiliation{%
 Ministry of Education Key Laboratory for Nonequilibrium Synthesis and Modulation of Condensed Matter, Shaanxi Province Key Laboratory of Advanced Functional Materials and Mesoscopic Physics, School of Physics, Xi’an Jiaotong University, Xi’an 710049, China 
}%
 
\author{Congzhe Yan}
\email{These authors contributed equally to this work}
\affiliation{%
 Division of Physics and Applied Physics, School of Physics and Mathematical Sciences, Nanyang Technological University, Singapore 637371, Singapore.} 
\affiliation{%
Department of Physics, University of Science and Technology of China, Hefei, Anhui, 230026, China} 

\author{Peiyuan Cui}
\affiliation{%
 Division of Physics and Applied Physics, School of Physics and Mathematical Sciences, Nanyang Technological University, Singapore 637371, Singapore.} 
%

\author{Zhijun Jiang}
\email{zjjiang@xjtu.edu.cn}
\affiliation{%
 Ministry of Education Key Laboratory for Nonequilibrium Synthesis and Modulation of Condensed Matter, Shaanxi Province Key Laboratory of Advanced Functional Materials and Mesoscopic Physics, School of Physics, Xi’an Jiaotong University, Xi’an 710049, China 
}%

\author{Yuanjun Jin}
\email{yuanjunjin@m.scnu.edu.cn}
\affiliation{%
 Guangdong Basic Research Center of Excellence for Structure and Fundamental Interactions of Matter,
 Guangdong Provincial Key Laboratory of Quantum Engineering and Quantum Materials,
 School of Physics, South China Normal University, Guangzhou 510006, China 
}%
 
\author{Guoqing Chang}
\email{guoqing.chang@ntu.edu.sg}
\affiliation{%
 Division of Physics and Applied Physics, School of Physics and Mathematical Sciences, Nanyang Technological University, Singapore 637371, Singapore.} 
%





\begin{abstract}
The magnons in antiferromagnetic insulators (AFIs) exhibit dual chirality, each carrying opposite spin angular momentum. However, in ${\cal PT}$-symmetric AFIs, the magnon bands remain degenerate. In this work, we introduce a new class of ${\cal PT}$-preserving AFIs in which the giant chiral splitting of magnons can be induced and controlled by an external electric field. Unlike conventional cases, such AFIs host a hidden dipole coupled to the antiferromagnetic order, which allows an external electric field to break the magnon sublattice symmetry and thereby largely lift the band degeneracy. Group theoretical analysis identifies the possible magnetic layer groups, while first-principles calculations and spin-wave theory reveal band splittings up to $\mbox{20}\,\mbox{meV}$ in $\mbox{Cr}_{2}\mbox{CCl}_{2}$ and $\mbox{Cr}_{2}\mbox{CBr}_{2}$ under the electric field of 0.2\,$\mbox{V/\AA}$, corresponding to an effective magnetic field of 200\,$\mbox{T}$. In addition, the electrically controlled magnon chiral splitting enables reversible switching of magnon-mediated spin currents. These findings open a new route toward nonvolatile spintronics based on magnons.
\end{abstract}

\maketitle


\textit{Introduction.}
The study of magnetic excitations provides critical insights into fundamental properties of bosons, as well as the applications of magnetic materials\,\cite{ chumak2014magnon, chumak2015magnon, onose2010observation, bader2010spintronics,pirro2021advances, mcclarty2022topological,lenk2011building, pirro2021advances, chen2025unconventional, chen_PRX_2018_8, mook_PRX_2021_11, wang2019nonreciprocity}. As elementary excitations in ordered magnets, magnons can carry spin angular momentum\,\cite{chumak2014magnon, chumak2015magnon, onose_science_2010_329, zhang_thermal_Physreport_2024_1070, cheng2016spin, zyuzin2016magnon} that can be harnessed for information encoding and processing\,\cite{qi2023giant, liu2021electric, kuzmenko2018switching, wan2025topological, ruckriegel2018bulk}. Unlike electrons, magnons can propagate over long distances in insulators without Joule heating\,\cite{chumak2014magnon, chumak2015magnon, onose2010observation, bader2010spintronics,pirro2021advances,  parsonnet2022nonvolatile, mcclarty2022topological, lenk2011building, wang2019nonreciprocity}; moreover, their wavelengths at a given frequency are orders of magnitude shorter than those of photons, making them ideal for wave-based spintronic devices at the nanoscale\,\cite{chumak2014magnon, chumak2015magnon}. 

\begin{figure*}
    	\centering
    	\includegraphics[scale=0.75]{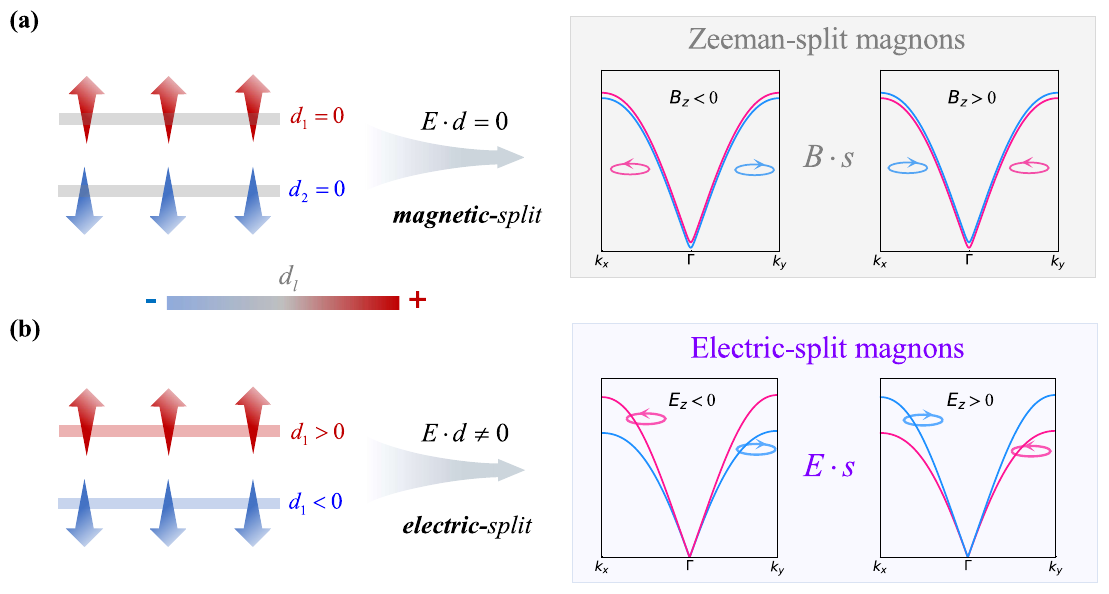}
    	\caption{The illustration of the magnons antiferromagnet with ${\cal PT}$ symmetry. (a) The degenerate magnons when the local dipole moment ($d_{l}$) is absent, where only the Zeeman field can split the bands. (b) The degenerate magnons undergo electric-field-induced splitting in the presence of $d_{l}$. Here, the red and blue circles represent the right- and left-handed chiralities of magnons, respectively. 
    \label{fig1}}
\end{figure*}

Ordered magnets are conventionally classified as ferromagnets or antiferromagnets, whose elementary spin excitations exhibit fundamentally distinct properties. In ferromagnets, magnons (ferromagnons) are polarized with a right-handed chirality\,\cite{chumak2015magnon}. In contrast, antiferromagnets host magnons (antiferromagnons) that naturally possess both left-handed and right-handed chiralities\,\cite{bader2010spintronics, baltz2018antiferromagnetic, jungwirth2016antiferromagnetic, rezende2019introduction}. At zero wave vector, the quadratic dispersion relation for ferromagnons  $\epsilon(\textbf{k})$\,$\propto$\,$\textbf{k}^{2}$, restricts their propagation frequency to the gigahertz\,(GHz) range, whereas antiferromagnons follow a linear dispersion relation, $\epsilon(\textbf{k})$\,$\propto$\,$\textbf{k}$, allowing them to operate in the terahertz\,(THz) range\,\cite{gomonay2018antiferromagnetic, rezende2019introduction}. However, two chiral modes are degenerate due to the ${\cal PT}$ symmetry in antiferromagnons, which prevents the generation of spin currents or net spin accumulation, limiting their use in spintronics. One way to overcome this limitation is to apply a strong magnetic field, though the resulting band splitting is quite small\,\cite{rezende2019introduction, baltz2018antiferromagnetic, neumann2022thermal}. Alternatively, the bond exchange associated with strong spin-orbit coupling (SOC) can split the magnon bands, but intense SOC typically leads to the non-conservation of spin angular momentum, which in turn increases dissipation effects\,\cite{gitgeatpong2017nonreciprocal, sato2019nonreciprocal, matsumoto2020nonreciprocal, hayami2022essential}.

Recent studies have established that, in altermagnets\,\cite{vsmejkal2022beyond, vsmejkal2022emerging, hayami2019momentum, xiao2024spin, chen2024enumeration, jiang2024enumeration, liu2022spin}, the magnon modes exhibit non-relativistic band splitting\,\cite{vsmejkal2023chiral, wu2025magnon, vsmejkal2023chiral, chen2025unconventional, zhang2025chiral, alaei2025origin, liu2024chiral, morano2025absence, sun2025observation, singh2025}. This splitting arises from the breaking of translational and inversion symmetry between spin-compensated sublattices, enabling the alternating chiral splitting. Meanwhile, the unique symmetry significantly restricts the variety of altermagnons, with only a few confirmed to date\,\cite{liu2024chiral, morano2025absence, sun2025observation, singh2025}. In addition, the symmetry constraint hinder external field manipulation. As a result, realizing magnons with robust splitting and external field tunability remains elusive. 

\setlength{\tabcolsep}{5pt} 
   \renewcommand\arraystretch{2} 
    \setlength{\tabcolsep}{1pt} 
\begin{table}[!h]
	\centering
    \renewcommand{\thetable}{\arabic{table}}
	\caption{Magnetic layer groups with polar magnetic WPs in high symmetry lattices enabling electric field-driven magnon band splitting in ${\cal PT}$-symmetric antiferromagnets. The complete classification is presented in Sec.\,IV of the SM\,\cite{Supplemental_Materials}.}
    \label{tab1}
  \begin{tabular}[b]{c|c|c}
   \hline
   \hline
  Point group  & \makecell[c]{Magnetic \\layer group}   & Magnetic WPs \\
  \hline
        \makecell[c]{$\bar{3}m$, $\bar{3}$ } & \makecell[c]{$p\bar{3}^{\prime}1m^{\prime}$ \\ $p\bar{3}^{\prime}1m$ \\$p\bar{3}^{\prime}m^{\prime}1$ \\ $p\bar{3}^{\prime}$} & \makecell[c]{2e ($3.m^{\prime}$), 4h ($3..$)\\ 4h ($3..$)\\ 2c ($3m^{\prime}.$), 2d ($3m^{\prime}.$)\\ 2c ($3..$), 2d ($3..$)} \\
        \hline
        \makecell[c]{$4/mmm$ } & \makecell[c]{$p4/m^{\prime}m^{\prime}m^{\prime}$, \\$p4^{\prime}/m^{\prime}m^{\prime}m$,\\$p4/n^{\prime}b^{\prime}m^{\prime}$, \\$p4/n^{\prime}bm$,\\$p4^{\prime}/n^{\prime}bm^{\prime}$,\\$p4/m^{\prime}b^{\prime}m^{\prime}$, \\$p4/m^{\prime}bm$,\\$p4^{\prime}/m^{\prime}bm^{\prime}$,\\$p4/n^{\prime}m^{\prime}m^{\prime}$, \\$p4^{\prime}/n^{\prime}m^{\prime}m$} & \makecell[c]{2g ($4m^{\prime}m^{\prime}$), 2h ($4m^{\prime}m^{\prime}$), 4i ($2m^{\prime}m^{\prime}$) \\ 4i ($2m^{\prime}m^{\prime}$) \\4g ($4..$), 4h ($2.m^{\prime}m^{\prime}$) \\ 4g ($4..$)\\ 4h ($2.m^{\prime}m^{\prime}$) \\4e ($4..$), 4f ($2.m^{\prime}m^{\prime}$) \\ 4e ($4..$)\\4f ($2.m^{\prime}m^{\prime}$) \\2c ($4mm$), 4f ($2m^{\prime}m^{\prime}$) \\ 4f ($2.m^{\prime}m^{\prime}$)} \\ 
        \hline
     \makecell[c]{$4/m$ } & \makecell[c]{$p4/m^{\prime}$,\\$p4^{\prime}/m^{\prime}$,\\$p4/n^{\prime}$,\\$p4^{\prime}/n^{\prime}$} & \makecell[c]{2g ($4..$), 2h ($4..$), 4i ($2..$)\\ 4i ($2..$) \\ 2c ($4..$), 4f ($2..$) \\ 4f ($2..$)} \\
     \hline
     	\makecell[c]{$6/mmm$ } & \makecell[c]{$p6/m^{\prime}m^{\prime}m^{\prime}$   \\ $p6^{\prime}/mmm^{\prime}$ \\ $p6^{\prime}/mm^{\prime}m$ \\}  & \makecell[c]{2e ($6m^{\prime}m^{\prime}$), 4h ($3m^{\prime}$),  6i ($2m^{\prime}m^{\prime}$) \\ 6j ($m2^{\prime}m^{\prime}$) , 12p ($m..$)\\ 4h ($3m^{\prime}$), 6l ($mm^{\prime}2^{\prime}$),  12p ($m..$)} \\
     \hline
     	\makecell[c]{$6/m$ } & \makecell[c]{$p6/m^{\prime}$,\\$p6^{\prime}/m$ } & \makecell[c]{2e ($6..$), 4h ($3..$), 6i ($2..$)  \\  4h ($3..$), 6j ($m..$)}\\ 
        \hline
        \hline
	\end{tabular}
\end{table}\label{tab1}
\renewcommand{\thetable}{\arabic{table}}

In this work, we identify a new class of ${\cal PT}$-preserving antiferromagnetic insulators (AFIs), where magnons exhibit giant chiral splitting solely induced by an electric field. Such AFIs feature a hidden dipole coupled to the antiferromagnetic order, allowing the electric field to break the sublattices of AFIs. This significantly modifies the intra-sublattice Heisenberg exchanges for opposite spins, lifting the magnon bands degeneracy. By enumerating the magnetic layer groups combined with density functional theory (DFT) calculations, we identify a series of candidate materials with magnon band splittings exceeding $\mbox{20}\,\mbox{meV}$. Importantly, this mechanism is independent of SOC, ensuring the robustness of the band splitting and the associated spin current at a finite temperature. 

\textit{Coupling between spin and hidden dipole.} To reveal the central role of hidden dipole in electric-split magnons from group theory, we first consider a $\mathcal{PT}$-symmetric antiferromagnets with magnetic layer group $G$. The $G$ can be decomposed into cosets as $G$\,=\,$H$\,+ \,$\mathcal{PT}H$\,+ $\cdots$\, + \,$R_nH$.
As $\mathcal{PT}$ is the coset representative, the subgroup $H$ breaks $\mathcal{P}$ symmetry and can therefore be polar, allowing for spontaneous polarization. The $H$ describes the site symmetry of a magnetic Wyckoff position (WP), which ensures that the magnetic atoms are electrically polarized\,\cite{fiebig2005revival, ni2025nonvolatile, shen2024}. By screening all $\mathcal{PT}$-symmetric groups, we identify all polarized magnetic WPs, see results in Tab.\,\ref{tab1} and Sec.\,IV of Supplemental Material (SM)\,\cite{Supplemental_Materials}. Particularly, as shown in Fig.\,\ref{fig1}(b), we introduce a hidden dipole vector $\boldsymbol{\lambda}(\mathbf{r}_1 - \mathbf{r}_2)$ ($\mathbf{r}_1$ and $\mathbf{r}_2$ are position vectors of two magnetic atoms) to capture the compensated polarization. The $\boldsymbol{\lambda}$ is polar under $\mathcal{P}$ symmetry: $\boldsymbol{\lambda}(\mathbf{r}_1 - \mathbf{r}_2) \rightarrow -\boldsymbol{\lambda}(\mathbf{r}_1 - \mathbf{r}_2)$ and its magnitude is proportional to the local dipole moment ($d_{l}$) and the distance between two magnetic atoms. The $\boldsymbol{\lambda}$\, implies an effective magnetoelectric\,(ME) coupling coefficient $\boldsymbol{\chi} \propto \boldsymbol{\lambda M}$ ($\boldsymbol{ M}$ is local magnetic moment), which is allowed under $\mathcal{PT}$ symmetry and yields a giant Zeeman-like field under external field $\boldsymbol{E}$, $B_\alpha = \chi_{\alpha \beta} E_\beta$.  Moreover, when the local dipole is coupled with the spin lattice, the polar distortion modifies the electronic wavefunctions and ligand fields in the material, generating an internal effective magnetic field\,\cite{zhao2022zeeman, tao2024layer, zhaozeeman2025}. Please note that in conventional $\mathcal{PT}$  antiferromagnet without hidden dipole moment ($\boldsymbol{\lambda}=0$), the magnon splitting is negligible under electric field\,\cite{ni2025magnon, du2025nonreciprocal}, and only an external magnetic field can induce a weak band splitting, as shown in Fig.\,\ref{fig1}(a). 

\begin{figure*}
\includegraphics[width=0.75\textwidth]{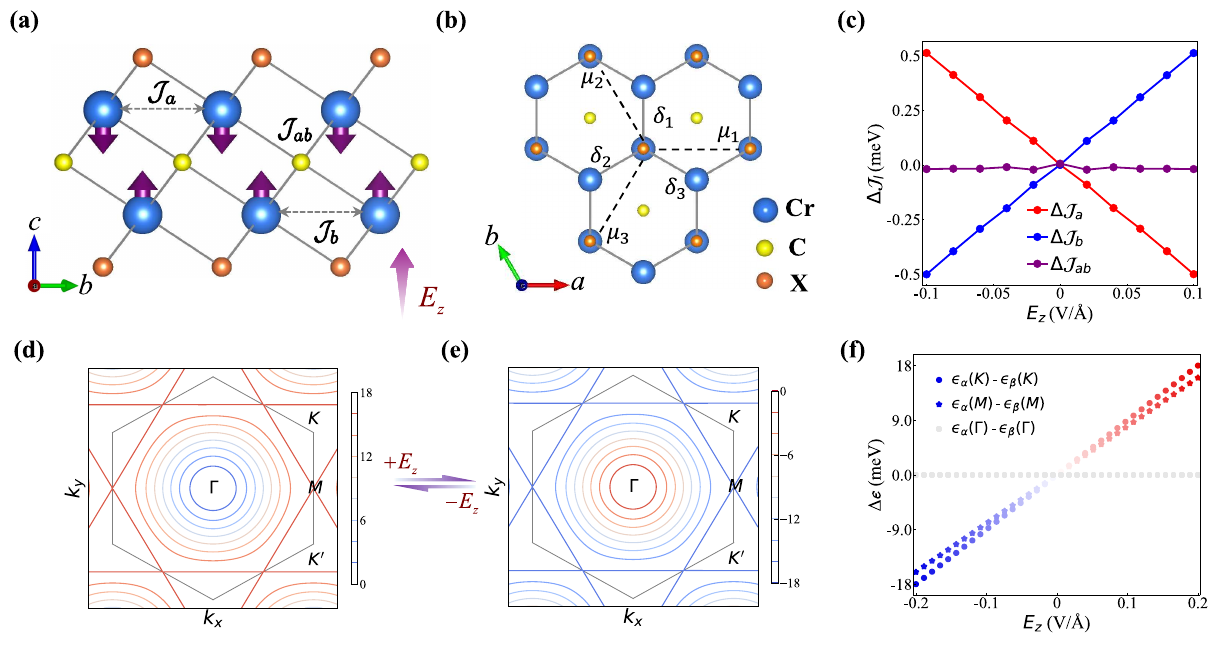}
\caption{\label{fig2}{(a)-(b) The side and top view of of the monolayer $\mbox{Cr}_{2}\mbox{C}X_{2}$ where $X$\,=\,$\mbox{F}$,\,$\mbox{Cl}$,\,$\mbox{Br}$. (c) The changes in the NN spin exchange ${\cal J}_{a}$, ${\cal J}_{b}$ and ${\cal J}_{ab}$ on $E_{z}$. Here, only the relative energy differences between the cases with and without the $E_{z}$ are presented, denoted as $\Delta{\cal J}_{a}$, $\Delta{\cal J}_{b}$ and $\Delta{\cal J}_{ab}$, respectively. The relative band splitting energies ($\epsilon_{\alpha}$\,$-$\,$\epsilon_{\beta}$) with (d), $E_{z}$\,$>$\,${0}$ and (e)\,$E_{z}$\,$<$\,${0}$. (f) The changes in the splitting strength at high symmetry points on $E_{z}$.}}
\end{figure*}

\textit{Material realization.}
Next, DFT calculations\,\cite{xiang2011predicting, xiang2013magnetic, ni2021giant} are performed to verify our proposed framework and demonstrate the physical origin of the magnon band splitting in monolayer $\mbox{Cr}_{2}\mbox{CF}_{2}$ under the electric field. Monolayer $\mbox{Cr}_{2}\mbox{CF}_{2}$ adopts the $p\bar{3}^{\prime}m^{\prime}1$ magnetic layer group, where Cr ions occupy WP 2d (0.333, 0.333, 0.5) with polar site group $3m^\prime$. In $\mbox{Cr}_{2}\mbox{CF}_{2}$, $\mbox{Cr}^{3+}$ ions adopt $S$\,=\,3/2 insulator states, forming the triangular magnetic lattice in each layer\,\cite{Half_chen, 9k4g-kjgf}. Obviously, as shown in Fig.\,\ref{fig2}(b), two $\mbox{Cr}$ layers with opposite layer polarization are stacked in an AB configuration, forming a honeycomb magnetic lattice. DFT calculations reveal that nearest-neighbor (NN) interlayer spin exchange (${\cal J}_{ab}$) is antiferromagnetic (AFM) coupling with $\mbox{8.3}$\,$\mbox{meV}$, and the intralayer spin exchanges (${\cal J}_{a}$ or ${\cal J}_{b}$) exhibits ferromagnetic coupling with $\mbox{9.5}$\,$\mbox{meV}$, allowing the N\'{e}el temperature ($T_{N}$) to exceed room temperature\,\cite{Supplemental_Materials}. When $E_{z}$\,$=$\,0, the local magnetic moments with opposite spins are equal ($|M_{a}|$\,$=$\,$|M_{b}|$), ensuring ${\cal J}_{a}$\,$=$\,${\cal J}_{b}$. 

Upon applying $E_{z}$, the ME coupling effect distributes the balance between the layers. Due to the nonzero hidden dipole vector $\boldsymbol{\lambda}$ which modifies the electric occupation and amplifies the Zeeman-like field, DFT calculations show that changes of the local magnetic moments reach a value of 0.004\,$\mu_{B}$ with $E_{z}$ of 0.2\,$\mbox{V/\AA}$ (see Fig.\,S1\,\cite{Supplemental_Materials}). The corresponding linear ME coupling coefficient is estimated to be -1.39$\times\mbox{10}^{-8}\,\mbox{s/m}$. Consequently, it significantly modifies the intralayer spin exchanges, as illustrated in Fig.\,\ref{fig2}(c), where
${\cal J}_{a}$\,-\,${\cal J}_{b}$\,=\,$-1\,\mbox{meV}$ for $E_{z}$\,$=$\,0.1\,$\mbox{V/\AA}$. As
the strength of $E_{z}$ increases, ${\cal J}_{a}$\,$-$\,${\cal J}_{b}$
exhibits a linear dependence on $E_{z}$, described by ${\cal J}_{a}$\,$-$\,${\cal J}_{b}$\,$=$\,$-10$\,$E_{z}$, indicating that electric field can reverse the sign of ${\cal J}_{a}$\,$-$\,${\cal J}_{b}$. In contrast, the changes in intralayer spin exchange ${\cal J}_{ab}$ and single-ion anisotropy (SIA) are negligible. As shown in Fig.\,\ref{fig2}\,(d), the values of ${\cal J}_{ab}$ remain approximately equal in the presence and absence of the electric field, indicating the N\'{e}el order remains unchanged under the electric fields. Notably, as shown in Tab.\,S1\,\cite{Supplemental_Materials}, for $E_{z}$\,$=$\,0.2\,$\mbox{V/\AA}$, ${\cal J}_{a}$\,$-$\,${\cal J}_{b}$\,$\approx$\,$0$ for $\mbox{Mn}\mbox{PS}_{3}$ and bilayer $\mbox{NiCl}_{2}$, primarily due to the absence of hidden dipole vector $\boldsymbol{\lambda}$.

\begin{figure*}
    	\centering
    	\includegraphics[scale=0.285]{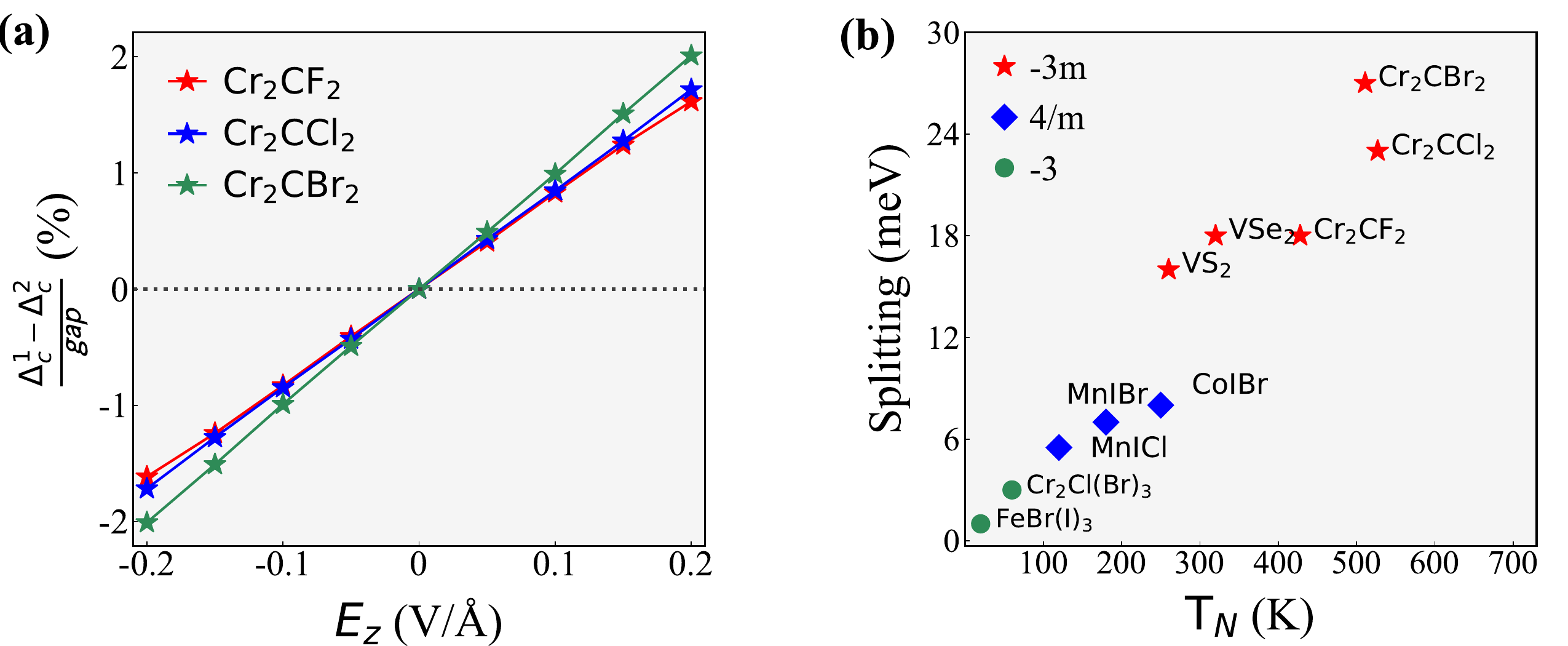}
    	\caption{(a) DFT calculated $E_{z}$ dependence of normalized $\frac{\Delta^{1}_{c}-\Delta^{2}_{c}}{gap}$ for $\mbox{Cr}_{2}\mbox{C}X_{2}$. Here, $\Delta^{1}_{c}$ and $\Delta^{2}_{c}$ refer to the crystal field splitting within the first layer and second layer of $\mbox{Cr}^{3+}$ ions, respectively, and $gap$ refers to the energy band gap. (b) The band splitting of varies material candidates with $E_{z}$\,$=$\,$0.2\,\mbox{V/\AA}$. For $-3$ point group such as bilayer $\mbox{Fe}_{2}\mbox{Br}_{3}\mbox{I}_{3}$, the magnetic ions occupy WP 2d (0.333, 0.667, 0.5) and  (0.0, 0.0), respectively.  In the $4/m$ point group, such as bilayer $\mbox{MnIBr}$, the magnetic ions occupy WP 2c (0.0, 0.0, 0.5). 
    \label{fig3}}
\end{figure*}

\textit{Magnon chiral splitting}. 
Building on the above DFT calculations and model analysis, we construct the effective spin Hamiltonian of monolayer $\mbox{Cr}_{2}\mbox{CF}_{2}$, expressed as 
\begin{equation}\label{Eq1}
\begin{split}
    {\cal \hat{H}}_{s} = & {\cal J}_{ab}\sum_{\langle i,j \rangle}{\cal S}_{i} \cdot {\cal S}_{j} + \sum_{\langle i,j \rangle} {\cal J}_{a}\left( {\cal S}_{i} \cdot {\cal S}_{j}\right)\\ &+ \sum_{\langle i,j \rangle} {\cal J}_{b}\left( {\cal S}_{i} \cdot {\cal S}_{j}\right) +  \sum_{i}{\cal K}\left({\cal S}^{z}_{i}\right)^{2}, 
\end{split}
\end{equation}
where the first term is NN interlayer spin exchanges and the second to third terms refer to the NN intralayer spin exchanges. The linear spin wave model in Eq.\,(\ref{Eq1}) can be solved by employing Holstein-Primakoff (HP) transformation\,\cite{holstein1940field}, ${\cal S}_{i,\uparrow}^{+} \approx \sqrt{2S} \hat{a}_{i}$, ${\cal S}_{i,\downarrow}^{-} \approx  \sqrt{{2S}}\hat{b}_{i}$, ${\cal S}_{i,\uparrow}^{z} = S - \hat{a}_{i}^{\dagger}\hat{a}_{i}$ and ${\cal S}_{i,\downarrow}^{z} = \hat{b}_{i}^{\dagger}\hat{b}_{i} - S$. Upon Fourier transformation, the Hamiltonian can be expressed in the basis $\psi^{\dagger}_{\textbf{k} } \equiv (\hat{a}^{\dagger}_{\textbf{k} }, \hat{b}_{-\textbf{k} } )$ as ${\cal {\hat{H}}} = \sum_{\textbf{k} }\psi^{\dagger}_{\textbf{k} } {\cal \hat{H}}_{\textbf{k} } \psi_{\textbf{k} }$. ${\cal \hat{H}}_{\textbf{k} }$ reads as, 
\begin{equation} {\label{Ham_k}}
\frac{{\cal \hat{H}}_{\textbf{k}}}{S} = 3{\cal J}_{ab} -2{\cal K} + f^{+}_{\textbf{k}} + { \begin{pmatrix}
    +f^{-}_{\textbf{k}}& \gamma_{\textbf{k}} \\
   \gamma^{\dagger}_{\textbf{k}} &  -f^{-}_{k} 
\end{pmatrix}}, 
\end{equation}
where 
\begin{equation}
\begin{split}
        \gamma_{\textbf{k} }  &= {\cal J}_{ab}\sum_{\textbf{k} ,\delta_{i}}e^{i\textbf{k}\cdot \boldsymbol{\delta}_{i}}, \\ f^{\pm}_{\textbf{k} } &= \frac{1}{2}\left({\cal J}_{a} \pm  {\cal J}_{b}\right)\sum_{\textbf{k} , \mu_{i}}2\mbox{cos}(\textbf{k}\cdot \boldsymbol{\mu}_{i})-6,
\end{split}
\end{equation}
with $\delta_{i}$ and $\mu_{i}$ being the NN and 2NN linking vectors between $\mbox{Cr}^{3+}$ ions as shown in Fig.\,\ref{fig2}(b). 

The Eq.\,(\ref{Ham_k}) can be diagonalized by the Bogoliubov transformation\,\cite{bogoljubov1958new, valatin1958comments, du2025nonreciprocal, ni2025magnon}, where the two physical solutions are given by 
\begin{equation}\label{eigen_AFM}
\begin{split}
{\epsilon}^{\pm}_{\textbf{k}}  = S\sqrt{\left[f^{+}_{\textbf{k} }+ 3{\cal J}_{ab} - 2{\cal K} \right]^{2} - \left|\gamma_{\textbf{k} }\right|^{2} } \pm f^{-}_{\textbf{k} }.
\end{split}
\end{equation}
Clearly, two magnon modes, denoted as $\beta$ and $\alpha$, are degenerate in the entire Brillouin zone when $E_{z}$\,=\,$0$, corresponding to ${\cal J}_{a}$\,=\,${\cal J}_{b}$. Since the Eq.\,(\ref{Ham_k}) is block diagonal, spin angular momentum along the $z$ direction, $\langle{s_{z}}\rangle$, is conserved. As a result, $\beta$ and $\alpha$ magnon modes carry opposite $\langle{s_{z}}\rangle$, with chirality $\pm{1}$, respectively. 

The nonzero ${\cal J}_{a}$\,$-$\,${\cal J}_{b}$ naturally breaks the degeneracy between $\alpha$ and $\beta$ modes. For $E_{z}$\,$>$\,$0$, the strength of ${\cal J}_{a}$ is larger than ${\cal J}_{b}$, lifts the $\alpha$ mode is high in energy than $\beta$ mode expect for $\Gamma$ point. This ordering is reversed when $E_{z}$\,$<$\,$0$. Compared with the magnon band splitting induced by an external magnetic field, the electric-field-induced band splitting exhibits anisotropy. Apart from the degeneracy at $\Gamma$ point, the magnitude of the splitting varies along the $\mbox{K}$-$\mbox{M}$ path. As shown in Figs.\,\ref{fig2}(d)-(f), the largest gap locate at the $\mbox{K}$-points, given as 
\begin{equation}\label{gap}
    \Delta_{\epsilon}(\mbox{K}) = 9\,\left|{\cal J}_{a}-{\cal J}_{b} \right|.
\end{equation}
Given that ${\cal J}_{a}$\,$-$\,${\cal J}_{b}$\,$=$\,$-10$\,$E_{z}$, $\Delta_{\epsilon}(\mbox{K})$ reaches at $\mbox{18}\,\mbox{meV}$ when $E_{z}$\,$=$\,$0.2\,\mbox{V/\AA}$ for $\mbox{Cr}_{2}\mbox {CF}_{2}$. The strength of the magnetic field required to induce the same magnitude of magnon band splitting can be estimated as 
\begin{equation}\label{Beff}
   {B}_{e} =\frac{\Delta_{\epsilon}(\mbox{K})}{g_{s}s_{z}}, 
\end{equation}
where $s_{z}$\,$=$\,$1$ and $g_{s}$ is the effective $g$ factor with assuming to 1. From Eq.\,(\ref{gap}) and Eq.\,(\ref{Beff}), the $E_{z}$ of 0.2\,$\mbox{V/\AA}$ produces a ${B}_{e}$ field as large as 155\,T. The thermal stability of magnon band splitting is confirmed by the renormalized spin wave theory (RSWT), with details provided in Section.\,III of the SM\,\cite{Supplemental_Materials, sourounis2024impact}. Unlike spin splitting induced by SOC, which leads to band instability due to spin relaxation and dissipation effects, the magnon band splitting induced by the electric field in $\mbox{Cr}_{2}\mbox{CF}_{2}$ with weaker SOC remains stable even above room temperature.

Additionally, we present the magnon band splitting of $\mbox{Cr}_{2}\mbox {CCl}_{2}$ and $\mbox{Cr}_{2}\mbox {CBr}_{2}$ under the electric field. The calculated ${\cal J}_{a}$\,-\,${\cal J}_{b}$ for $\mbox{Cr}_{2}\mbox{CCl}_{2}$ and $\mbox{Cr}_{2}\mbox {CBr}_{2}$ at $E_{z}$\,$=$\,$0.2\,\mbox{V/\AA}$ are 2.6\,$\mbox{meV}$ and 3.1\,$\mbox{meV}$, respectively, corresponding to $\Delta_{\epsilon}(\mbox{K})$ of $\mbox{23}\,\mbox{meV}$ and $\mbox{27}\,\mbox{meV}$. These values are significantly larger than those observed in $\mbox{Cr}_{2}\mbox {CF}_{2}$. Notably, the linear ME coupling coefficient of $\mbox{Cr}_{2}\mbox {CCl}_{2}$ and $\mbox{Cr}_{2}\mbox {CBr}_{2}$ are -1.38$\times\mbox{10}^{-8}s/m$ and -1.19$\times\mbox{10}^{-8}s/m$, respectively, both smaller than the strength observed in $\mbox{Cr}_{2}\mbox {CF}_{2}$. To elucidate this discrepancy, we employ the multi-orbital model to deduce the form of the spin exchanges\,\cite{winkler2003spin}, as ${\cal J}$\,$\propto$\, $1/\Delta_{c}$, where $\Delta_{c}$ refers to the energy gap between spin-empty and spin-occupied states. Consequently, the magnitude of ${\cal J}_{a}$\,-\,${\cal J}_{b}$ can be quantitatively characterized by the layer-resolved normalized gap difference $\frac{\Delta^{1}_{c}-\Delta^{2}_{c}}{gap}$. Fig.\,\ref{fig3}(a) shows the $E_{z}$ dependence of normalized $\frac{\Delta^{1}_{c}-\Delta^{2}_{c}}{gap}$ for $\mbox{Cr}_{2}\mbox{C}X_{2}$. Clearly, the $\mbox{Cr}_{2}\mbox{CBr}_{2}$ responds stronger value than that of the $\mbox{Cr}_{2}\mbox{CCl}_{2}$ and $\mbox{Cr}_{2}\mbox{CF}_{2}$, which explains the distinct electric-field dependence of their intralayer spin exchange. Guided by symmetry analysis and high-throughput calculations, we systematically examine magnon band splitting in antiferromagnets with hexagonal and tetragonal crystal symmetries. As shown in Fig.\,\ref{fig3}(b), the magnitude of magnon band splitting is mainly determined by the strength of ME related to the hidden dipole moment, and modified by $\text{T}_{N}$ and energy band gap.


\newcommand{\opalpha}{\alpha\vphantom{\beta}}
\newcommand{\opbeta}{\beta\vphantom{\alpha}}
\begin{figure}
    	\centering
    	\includegraphics[scale=0.16]{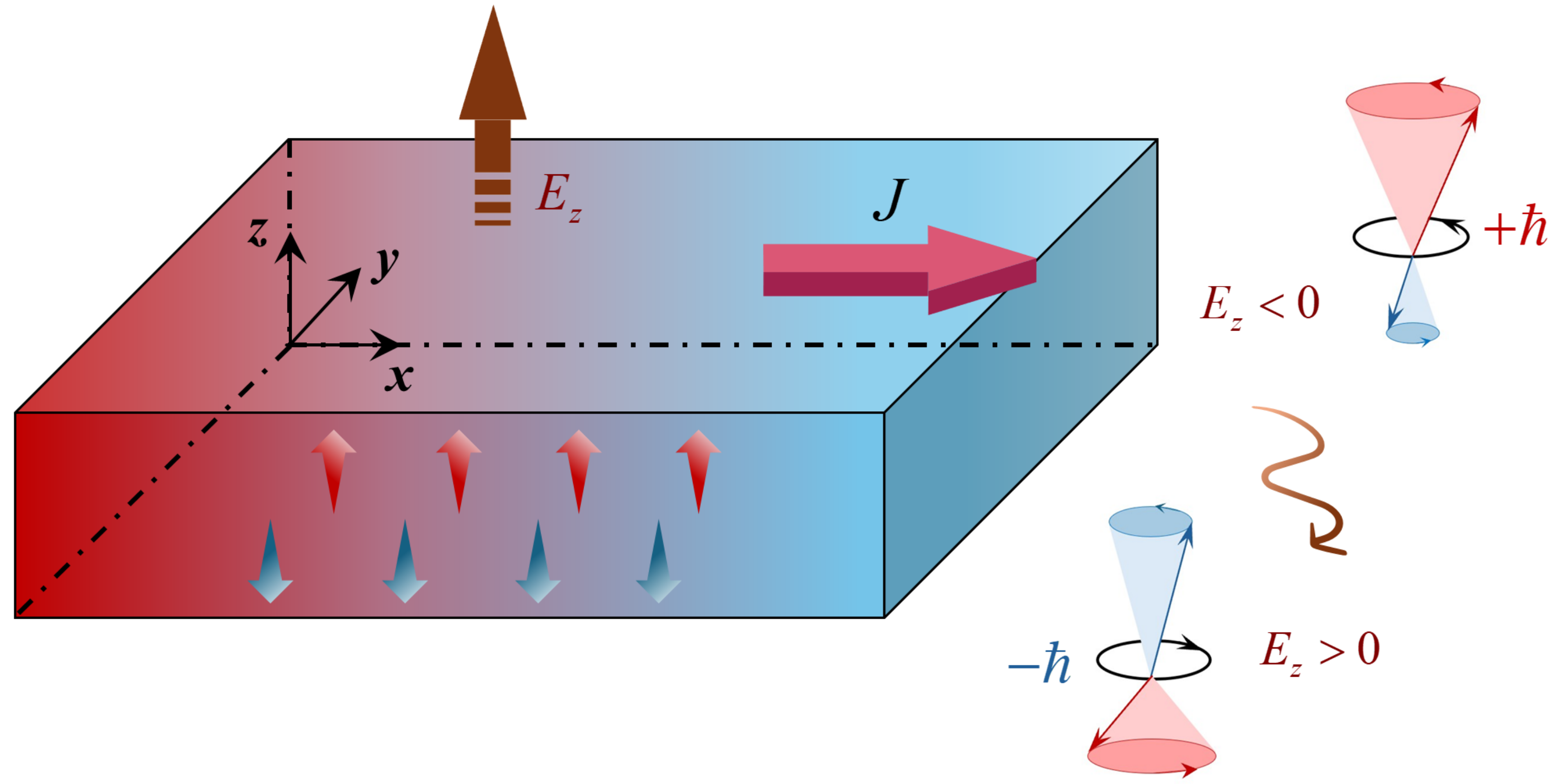}
    	\caption{Illustration of electric switchable spin currents mediated by the magnons in antiferromagnetic insulators. Here, the transition from red to blue represents the distribution of the applied temperature gradient.
    \label{fig4}}
\end{figure} 

\textit{Electric-switchable spin currents}. The giant chiral splitting of magnons naturally leads to a spin current driven by the applied temperature gradient $\nabla{T}$ (see Fig.\,\ref{fig4}), known as the spin Seebeck effect\,\cite{cui2023efficient, wu2025magnon, onose2010observation, cheng2016spin, zyuzin2016magnon}. The thermal spin conductivity tensor $\sigma_{mn}$ is defined by $\langle{\bm{J}}\rangle_m$\,$=$\,$-\sigma_{mn}(\nabla{T})_n$, where $m,n$\,$\in$\,$\{x,y\}$, $\langle{\bm{J}}\rangle$ is the spin current density. Based on the Kubo formula\,\cite{Supplemental_Materials}, the thermal spin conductivity tensor $\sigma_{mn}$ is given as 
\begin{equation}\label{SC}
\begin{split}
\sigma_{mn}={\sum_{\lambda=\pm}\sum_{\textbf{k}}}{\frac{-\lambda\tau_{0}}{\hbar{{k} _B}VT^{2}}}{{\bm v}_{m,\textbf{k}}^{\lambda}}{{\bm v}_{n,\textbf{k}}^{\lambda}}\frac{{\epsilon}_{\textbf{k}}^{\lambda} e^{{\epsilon}_{\textbf{k}}^{\lambda}/{{k}_B}T}}{\left(e^{{\epsilon}_{{\textbf{k} }}^{\lambda}/{k_B}T}-1\right)^2} 
\end{split}
\end{equation}
where ${\bm v}^{\lambda}_{\textbf{k}}$\,$=$\,$\frac{{\partial \epsilon}^{\lambda}_{\textbf{k} }}{{\hbar\partial{\textbf{k} }}}$, $V$ is the volume and $\tau_0$ is the average magnon lifetime. For $\mbox{Cr}_{2}\mbox{C}X_{2}$ with $C_{3z}$ symmetry, the thermal spin conductivity tensor is anisotropic and could be simplified into a scalar $\sigma$\,$=$\,$\sigma_{xx}$\,$=$\,$\sigma_{yy}$ as $\sigma_{xy}$\,$=$\,$\sigma_{yx}$\,$=$\,$0$.

The linear dependence of thermal spin conductivity on applied electric field is shown in Fig.\,S2\,\cite{Supplemental_Materials}. For $E_{z}$\,$=$\,$0.1$\,$\mbox{V/\AA}$, $\sigma$\,$=$\,$-0.102$\,$\mbox{meV/K}$ at $T$\,$=$\,$300$\,$\mbox{K}$, which is three orders in magnitude larger than that induced by Berry curvature in monolayer and bilayer antiferromagnets\,\cite{cheng2016spin, zyuzin2016magnon}. As for $E_{z}$\,$<$\,$0$, the bandwidth $\epsilon_{\beta}(\mbox{K})$\,$-$\,$\epsilon_{\beta}(\Gamma$) exceeds $\epsilon_{\alpha}(\mbox{K})$\,$-$\,$\epsilon_{\alpha}(\Gamma)$, and consequently the effective velocity ${\bm v}_{\textbf{k} }^{\opalpha}$ exceeds ${\bm v}_{\textbf{k} }^{\opbeta}$. As a result, the contribution to the spin conductivity from the $\beta$ band is larger than that from the $\alpha$ band, leading to a positive spin conductivity. Since the $\alpha$ and $\beta$ magnon modes carry opposite $\langle{s_{z}}\rangle$, the electric field switches the sign of the spin current while also altering the intrinsic spin angular momentum information encoded, as shown in Fig.\,\ref{fig4}. This opens up new avenues for the electric-field-controlled spin currents.

\textit{Summary and discussion.} 
In summary, we have revealed a class of ${\cal PT}$-preserving AFIs that exhibit giant magnon band splitting under an electric field. Such pronounced electric-field-induced magnon splitting originate from a hidden dipole moment that amplifies the ME effect. Through DFT calculations combined with linear spin wave theory, we identified a series of candidate materials with magnon band splittings exceeding $\mbox{20}\,\mbox{meV}$. These candidates feature large band splitting and exceptional thermal stability, promising a design strategy for electrically controlled magnetism.

Moreover, the magnon band splitting induced by the electric field breaks the antiferromagnetic lattice symmetry but is independent of the SOC effect. As a result, the split magnon bands still conserve spin angular momentum. This behavior is distinct from the splitting caused by SOC, which leads to magnon dissipation; consequently, the electric-field-driven magnon splitting exhibits exhibits a longer lifetime and enhanced robustness. The associated thermal spin conductivity can be reversed by an electric field and and is significantly larger than that arising from magnon Berry curvature. Notably, all candidate materials considered here are insulators, which effectively prevents the band damping\,\cite{chumak2014magnon, chumak2015magnon, poelchen2023long, vsmejkal2023chiral}. Our work provides a new way toward nonvolatile antiferromagnetic spintronics based on the magnons.

\textit{Acknowledgments.} The authors thank Prof.\,Jian Zhou for helpful discussions. This work is supported by the National Natural Science Foundation of China (Grant No.\,12374092), Natural Science Basic Research Program of Shaanxi (Program No.\,2023-JC-YB-017), Shaanxi Fundamental Science Research Project for Mathematics and Physics (Grant No.\,22JSQ013), “Young Talent Support Plan” of Xi'an Jiaotong University, and the Xiaomi Young Talents Program. G.\,Q.\,Chang thanks grants from the Singapore National Research Foundation Fellowship Award (NRF-NRFF13-2021-0010) and  Nanyang Technological University startup grant (NTUSUG). Y.\,Jin thanks to the Guangdong Provincial Quantum Science Strategic Initiative (Grant No. GDZX2401002), the Natural Science Foundation of China (Grant No. 12404181).

\textit{Data availability}. All data are available from the authors upon reasonable request.

\nocite{*}
\bibliography{main}

\end{document}